\documentclass[letterpaper]{article}

\usepackage[T1]{fontenc}

\usepackage{geometry}
\usepackage{setspace}

\usepackage[
  style = chem-acs,
  maxnames = 15,        % Name up to 15 authors before et al.
  articletitle = true,   % Include the article title in citations
  doi = true 
]{biblatex}

\usepackage{graphicx}
\usepackage{float}
\newfloat{scheme}{htbp}{los}
\floatname{scheme}{Scheme}
\floatname{chart}{Chart}
\newfloat{graph}{htbp}{loh}
\usepackage{chemformula} % Formulas using \ch{}
\usepackage[version = 4]{mhchem} % Formulas using \ce{}
\usepackage{graphicx}
\usepackage{float}
\usepackage{subfig}
\usepackage[section]{placeins}
\usepackage{soul}
\usepackage{authblk}
\author[1]{Abrar Faiyad}
\author[1]{Ashlie Martini*}
\affil[1]{University of California Merced, Merced, California 95343, United States}

\title{Machine Learning Interatomic Potentials Enable Molecular Dynamics Simulations of Doped MoS$_2$}
\date{*Email: amartini@ucmerced.edu}

\begin{document}

\maketitle

\begin{abstract}
    Dopants can tune the performance of MoS$_2$ in various applications, but use of molecular dynamics simulations for doped MoS$_2$ materials discovery is limited by the lack of multi-dopant interatomic potentials.
    Universal machine learning interatomic potentials (MLIPs) could be a solution, but the accuracy of these potentials must first be evaluated. Here, we evaluate the accuracy of a recently developed MLIP, META's Universal Model for Atoms (UMA), for 25 different MoS$_2$ dopants spanning metals, non-metals, and transition metals in Mo substitution, S substitution, and intercalated positions by benchmarking the MLIP-predicted formation energy and the dopant-induced structural change against density functional theory calculations. 
    The computational framework for MLIP validation and simulations are described in detail and the source code is made open source.
    The MLIP is then demonstrated by performing heating-cooling simulations of MoS$_2$ supercells with all 25 dopants. These simulations capture complex phenomena including dopant clustering, MoS$_2$ layer fracturing, interlayer diffusion, and chemical compound formation at orders-of-magnitude reduced computational cost compared to density functional theory. This work provides a computational workflow for application-oriented design of doped-MoS$_2$, enabling high-throughput screening of dopant candidates and optimization of compositions for targeted tribological, electronic, and optoelectronic performance. 
\end{abstract}

\begin{figure}[ht]
    \centering
    \includegraphics[width=0.7\textwidth]{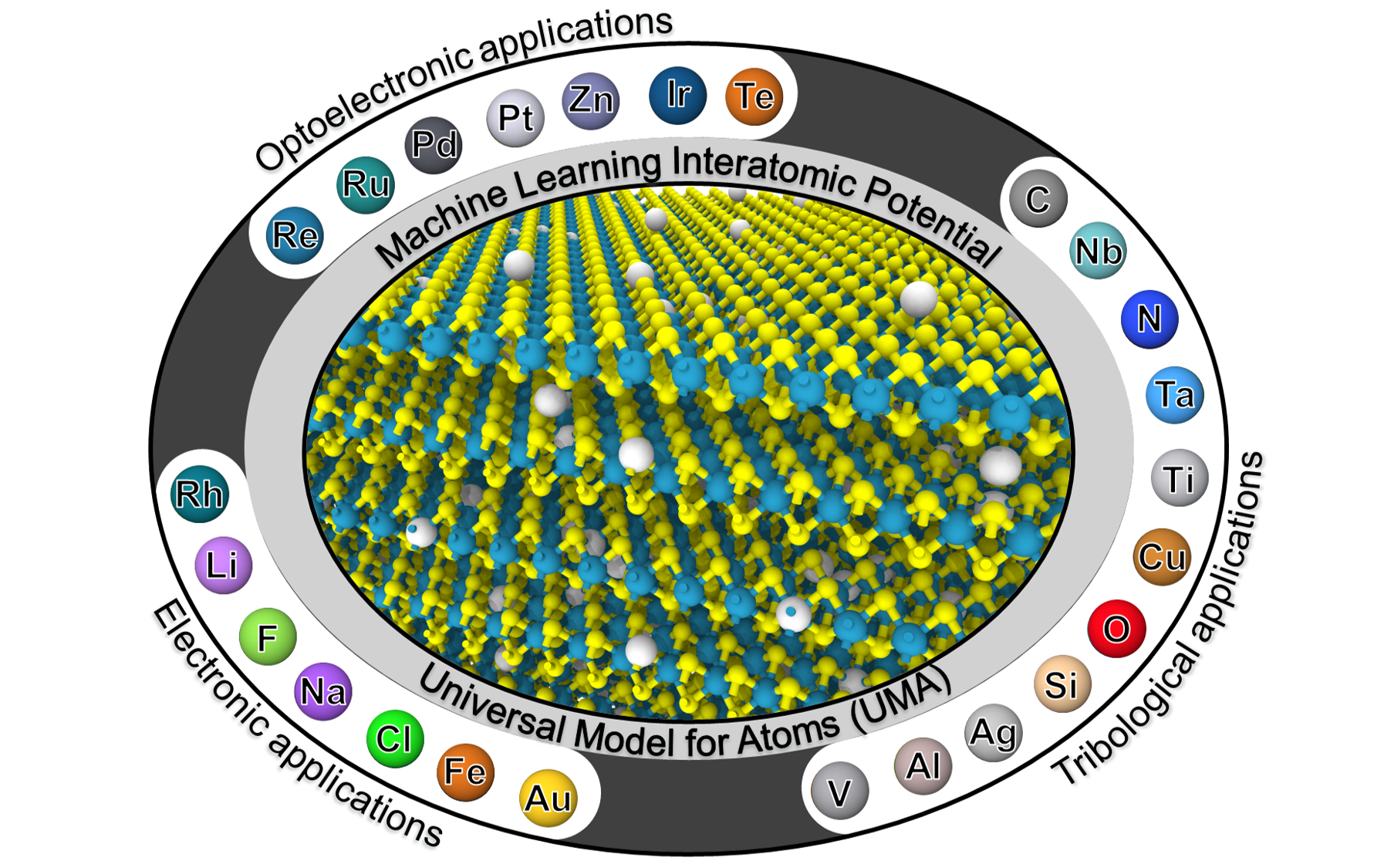}
    \label{fig:ToC}
\end{figure}
\FloatBarrier

\section*{Keywords}

Machine learning, Molecular dynamics, MoS2, Two-dimensional materials, Doping, MLIP, High-throughput screening

%\section*{Abbreviations}

%Some journals require a list of abbreviations: these normally should be given
%immediately after the keyswords (if required).

%%%%%%%%%%%%%%%%%%%%%%%%%%%%%%%%%%%%%%%%%%%%%%%%%%%%%%%%%%%%%%%%%%%%%
%% Start the main part of the manuscript here.
%%%%%%%%%%%%%%%%%%%%%%%%%%%%%%%%%%%%%%%%%%%%%%%%%%%%%%%%%%%%%%%%%%%%%
\section{Introduction}

\label{sec:introduction:2D Materials revolution}
% Properties
Transition metal dichalcogenides (TMDs) have fundamentally transformed materials science and engineering over the past two decades, giving rise to a new era of atomic-scale engineering \cite{Kumar2022-zl,Ozugur-Uysal2022-eq}. Among the expansive family of TMDs, molybdenum disulfide (MoS$_2$) stands out due to its unique combination of mechanical, electronic, and optical properties that position it at the forefront of next-generation technological applications \cite{Alharthi2024-zb,Verma2024-ww,Zeimpekis2024-of}. Unlike its semi-metallic counterpart graphene, which lacks an intrinsic bandgap, MoS$_2$ exhibits a tunable bandgap that transitions from indirect (1.2 eV) in bulk form to direct (1.9 eV) in monolayer configuration due to quantum confinement effects \cite{Lu2014-mq, Gerbi2025-dh, Lembke2015-ui}. 
MoS$_2$ also has mechanical strength comparable to steel while possessing good chemical stability and processability \cite{Hemanth2021-bp, Pang2020-rz}. Further, MoS$_2$ is a layered material. The layers of MoS$_2$ are held together by weak van der Waals forces which allows for easy exfoliation to monolayer and facilitates low resistance to sliding between layers \cite{Vazirisereshk2019-rc}.

% Applications
These unique properties have made MoS$_2$ a prime candidate for various applications. For example, The mechanical strength and low-resistance to sliding of MoS$_2$ result in low friction and wear, essential for tribological applications \cite{Faiyad2025-le, Vazirisereshk2019-rc}. In electronics, field-effect transistors (FETs) based on single-layer MoS$_2$ exhibit very high current on/off ratios (exceeding 10$^8$) and good switching characteristics. \cite{Lembke2015-ui}. MoS$_2$ is also used for energy storage \cite{Raza2025-dc,Zeimpekis2024-of} and as catalysts in hydrogen evolution reactions and oxygen evolution reactions \cite{Hemanth2021-bp, Mphuthi2022-pf}. Optoelectronic applications leverage the direct bandgap of MoS$_2$ monolayer to enable effective absorption and emission across the visible spectrum \cite{Lu2014-mq, Timpel2021-ij, Gerbi2025-dh, Wang2018-or}. Photodetectors based on monolayer and few-layer MoS$_2$ exhibit ultra fast photo response with carrier extraction occurring on the femtosecond to millisecond timescale, high photoresponsivity and external quantum efficiencies reaching up to 7~\% \cite{Wang2018-or}. Light-emitting diodes and related heterostructures show strong photoluminesence at 665~nm with quantum yields approximately three orders of magnitude higher than bulk MoS$_2$ and direct bandgap emission suitable for display applications \cite{Gerbi2025-dh}.

\label{sec:introduction:tuning properties via dopants}
Impurity atoms can be intentionally introduced into the MoS$_2$ crystal lattice, i.e., dopants, to enhance its physical, chemical or electronic properties for the discussed applications.
When Mo is substituted with Re, Ta, V and Tc, or Li, Na and K are absorbed on the surface, it creates an n-type semiconductor, increasing electron concentration and electrical conductivity \cite{Bae2024-cl}. Alternatively, doping with Ag, Au, Cu, C, P  Nb, N, As and Sb in different sites create a p-type semiconductor, allowing it to conduct electricity through positive charge carriers \cite{Bae2024-cl, Li2022-it, Kamruzzaman2021-eu}.
Gas dopants such as molecular O, NO$_2$ and NH$_3$ can reversibly modify the electronic properties of MoS$_2$, providing opportunities for tunable sensors and adaptive electronic devices \cite{Koci2023-hu}. 
Sb doping has been explored for thermoelectric applications, where the heavy atom mass contributes to reduced thermal conductivity while maintaining electrical conductivity \cite{Wu2024-tu}. 
Transition metal dopants including Co, Ni, Ru, and Fe, as well as non-metals like N, enhance the catalytic activity for hydrogen evolution reaction \cite{Jiang2021-br, Nolan2023-ui, Wang2021-pw}.
Finally, a wide variety of dopants have shown promise in tribological coatings with various mechanisms proposed to explain the benefit of the dopant.
It has been suggested that Cr, Ti, Zr, Ni, and Co doping increases density and hardness of MoS$_2$ coatings, soft metals like Ag, Au, Pb, and Sb act as a low-shear phase to decrease friction and wear, and C-based dopants add load-bearing capacity and improve oxidation resistance \cite{Ahmed2025-kc}. 
However, most experimental studies necessarily focus on just one or a few dopants due to the time and expense of material synthesis and characterization.
These practical constraints limit the extent to which the parameter space, i.e., dopant element and concentration, can be explore experimentally.

\label{sec:introduction:Need for Atom-Level Insight}
Atomistic simulations overcome this limitation and have been used to study the effect of dopants on MoS$_2$ properties and performance in target applications.
% DFT
Most atomistic studies performed on doped MoS$_2$ used density functional theory (DFT). DFT provides quantum-mechanically accurate descriptions of electronic structure, energetics, and chemical bonding, and it has been extensively used to study various chemical and electrical properties of MoS$_2$ \cite{Dolui2013-zl, Bae2024-cl, Gerbi2025-dh, Wang2023-tt, Vazirisereshk2019-rc, Li2022-it, Ramsey2025-vz, Raza2025-dc}. However, the accuracy of DFT calculations comes as the cost of computational efficiency such that they are limited to tens to hundreds of atoms and timescales on the order of femtoseconds. 
This is an issue for modeling doped MoS$_2$ since extended simulation sizes and timescales are needed to capture collective phenomenon such as dopant diffusion, clustering, interface formation, phase transition and long range strain fields - processes that fundamentally govern the functional performance of the material \cite{Romero-Garcia2024-zq, Mohammadtabar2023-eu, Rai2024-fj}. 

\label{sec:introduction:MD with non-ML potential}
To address the computational limitations of DFT, researchers have turned to molecular dynamics (MD) simulations using empirical approximations, or potentials. These approximations enable simulations of thousands to millions of atoms and time scales orders of magnitudes longer than DFT. For doped MoS$_2$ specifically, one study performed MD simulations of Cr-doped MoS$_2$ using newly developed CHARMM and CVFF potential parameters and showed that Cr doping significantly affects structural stability and increases hydrophobicity \cite{Xing2022-sm}. Other research teams developed ReaxFF parameters for Ni-doped MoS$_2$ and Ti-doped MoS$_2$, enabling reactive simulations of phase transitions from amorphous to crystalline structures during annealing \cite{Mohammadtabar2023-eu, Mao2022-ah}. The Ni-doped potential was later used to study the effects of dopant composition on MoS$_2$ crystallization \cite{Romero-Garcia2024-zq}. More recently, a ReaxFF potential was developed to study C doped MoS$_2$ systems \cite{Ponomarev2024-fw}. However, empirical potentials suffer from fundamental limitations in transferability and accuracy, particularly when applied to systems or conditions significantly different from those used in their parameterization. The inherent trade-off between computational efficiency and chemical accuracy in empirical potentials becomes particularly problematic for doped systems where electronic effects determine dopant stability and property modifications.

\label{sec:introduction:MD with ML potential}
To address this issue, machine learning interatomic potentials (MLIP) are rapidly emerging as transformative approaches that bridge the accuracy-efficiency gap between DFT and MD. 
Researchers have started using MLIPs to study TMDs \cite{Zhao2024-hn, Kety2024-zx,Flototto2025-fn, Siddiqui2025-tv, Siddiqui2024-ar, Liu2024-lw, Gu2019-zz, Marmolejo-Tejada2022-zs, Wang2019-dm}. 
Most of these studies focused on developing MLIPs for specific material systems, particularly, bilayers, heterostructures, and alloys of MoS$_2$, MoSe$_2$, WS$_2$, and WSe$_2$.
MLIPs were used to simulate the growth process of MoS$_2$/WS$_2$ van der Waals heterostructures\cite{Zhao2024-hn}, the formation of amorphous MoS$_2$ through melting-quenching\cite{Kety2024-zx}, sulfur vacancy dynamics in MoS$_2$ monolayer\cite{Flototto2025-fn}, and twisted bilayers of MX$_2$ (where M=Mo/W and X=S/Se)\cite{Siddiqui2025-tv}. 
MLIP-based simulations have reproduced the vibrational spectrum of Mo$_{1-x}$W$_x$S$_{2-2y}$Se$_{2y}$ quaternary alloy TMDs\cite{Siddiqui2024-ar}, determined the lattice thermal conductivity of monolayer MoS$_{2(1-x)}$Se$_{2x}$\cite{Gu2019-zz} and monolayer MoS$_{2(1-x)}$Se$_{2x}$\cite{Marmolejo-Tejada2022-zs}, and calculated the mechanical properties of WS$_2$ monolayer\cite{Wang2019-dm} and MoS$_2$/WS$_2$ alloys\cite{Liu2024-lw}. All MLIPs that have been trained for TMDs are either for pristine systems or alloys of TMDs.

Importantly, a fundamental limitation of both reactive empirical potentials and standard MLIPs is that they are generally only trained for only one or a few specific materials systems due to computational limitations.
As a result, they cannot be applied within a single unified framework to model MoS$_2$ with varied dopant chemistries and concentrations.
This lack of a unified framework not only restricts the ability to explore the full design space needed for materials exploration and optimization, but also inhibits the use of MD for addressing fundamental scientific questions about how different dopants modify the material properties of MoS$_2$ at an atomic scale across compositions.

% Universal MLIPs and UMA 
Recently, universal MLIPs trained on large DFT datasets containing elements and compounds spanning the periodic table have been developed. One such universal MLIP is META's Universal Model for Atoms (UMA), a family of machine-learning interatomic potential trained on half a billion unique 3D atomic structures spanning molecules, materials, and catalysts \cite{Wood2025-it}. Two sizes of the models are publicly available, the UMA small and UMA medium variants. UMA medium has 1.4 billion total model parameters and has a slower inference speed compared to UMA small which has 150 million total model parameters. Both models utilize a novel mixture-of-linear-experts architecture, activating only a fraction of parameters for each atomic structure, which greatly enhances computational efficiency \cite{Wood2025-it}. 
Both models are open-source and perform comparably or better than task-specific potentials, enabling large-scale, high-fidelity MD simulations across diverse chemical domains without finetuning. 
Thus, these universal MLIPs could complement experiments by enabling systematic studies of dopant incorporation, concentration-dependent behavior, and the influence of deposition parameters or methods.
They could also support the development of high throughput approaches for exploring dopant compositions and tuning target properties. But, before that, their accuracy specifically for doped MoS{$_2$} must be evaluated.
%Here, we evaluate META's UMA MLIP for doped MoS$_2$.

\label{sec:introduction:Scope of this work}
This study presents a comprehensive validation of formation energy and local structure of 25 dopants in MoS$_2$ systems using universal MLIPs. We benchmark UMA small and UMA medium against reference DFT calculations, quantifying their accuracy and reliability for predicting energetics and relaxed structures of three dopant locations in MoS$_2$ for each dopant. 
The accuracy of the MLIP is evaluated per dopant and location in terms of formation energy and, for the substituted dopants, nearest neighbor distance.
The code used for validation is available on github \cite{FaiyadUnknown-dv}.
To demonstrate the application of the MLIPs, we use them to run heating and cooling molecular dynamics with all 25 dopants.
Calculated dopant-dependent densities and atom mobility and visual analysis of the simulation trajectories show the potential for these MLIPs to capture complex phenomena including dopant clustering, MoS$_2$ layer fracturing, interlayer diffusion, and chemical compound formation at orders-of-magnitude reduced computational cost compared to density functional theory.
This work thus lays the foundation of future investigations of doped MoS$_2$ and other TMDs using universal MLIPs.

\section{Methods}
\label{sec:methods}

\subsection{Software and models}
We use Python and the Atomic Simulation Environment (ASE)~\cite{Hjorth-Larsen2017-zj} for optimization and job control. For machine-learning molecular dynamics (MLMD), FAIRChemCalculator (2.3.0) ~\cite{Wood2025-it,UnknownUnknown-zq} with UMA models- UMA small(\texttt{uma-sm-1p1}) and UMA medium (\texttt{uma-m-1p1}), and task OMAT are used. For the DFT calculations, \textsc{Quantum~ESPRESSO} (QE) with the PBE functional~\cite{Perdew1996-ij} and PSLibrary~1.0.0 pseudopotentials~\cite{Dal-Corso2014-lu}, are used following standard QE references~\cite{Giannozzi2009-sf,Giannozzi2017-du}. No spin–orbit coupling (SOC) is enabled. Ovito is used for visualization \cite{Stukowski2009-cu}.

In both MLMD and DFT, the atomic positions are optimized with the Broyden Fletcher Goldfarb Shann (BFGS) algorithm with fixed cells. A force convergence threshold of $5\times10^{-3}$~eV/\AA{} is used. For MLMD, ASE is used to relax each input geometry with BFGS to the common force threshold. We record the final total energy $E_{\mathrm{tot}}$ for formation-energy analysis. All MLMD calculations are performed on either Nvidia A100s or L40 GPUs. The ASE calculators are assigned per-structure in a single-stage relax-and-evaluate workflow~\cite{UnknownUnknown-zq}.

The DFT calculations use a fixed cell relaxation method with spin polarization and no SOC. We use the PBE functional from PSLibrary~1.0.0 PAW/USPP files. We test gamma point calculation, $4\times4\times4$ and $6\times6\times4$ Monkhorst–Pack mesh to test energy convergence, Fermi–Dirac smearing, and plane-wave cutoffs of $124/843$ Ry for wavefunctions/charge density, respectively. The plane wave cuttoffs are selected from the maximum lowest suggested cutoffs multiplied by $1.2$  in the pseudopotentials throughout all dopants. Electronic thresholds are $10^{-4}$ with \texttt{rmm-diis} diagonalization. All relaxations use BFGS with the same $5\times10^{-3}$~eV/\AA{} force target as MLMD. 

\subsection{Dopant set and formation-energy formalism}
\begin{figure*}[ht]
    \centering
    \subfloat[S substitution]{%
        \includegraphics[width=0.33\textwidth]{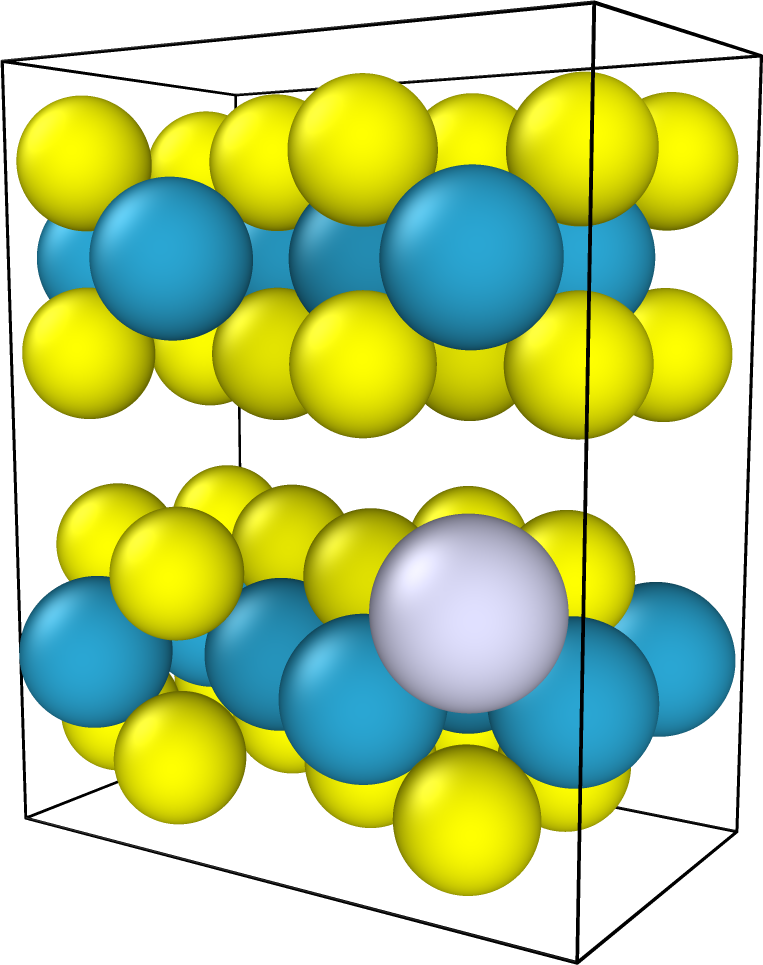}%
        \label{fig:s_sub}%
    }
    \hfill
    \subfloat[Mo substitution]{%
        \includegraphics[width=0.33\textwidth]{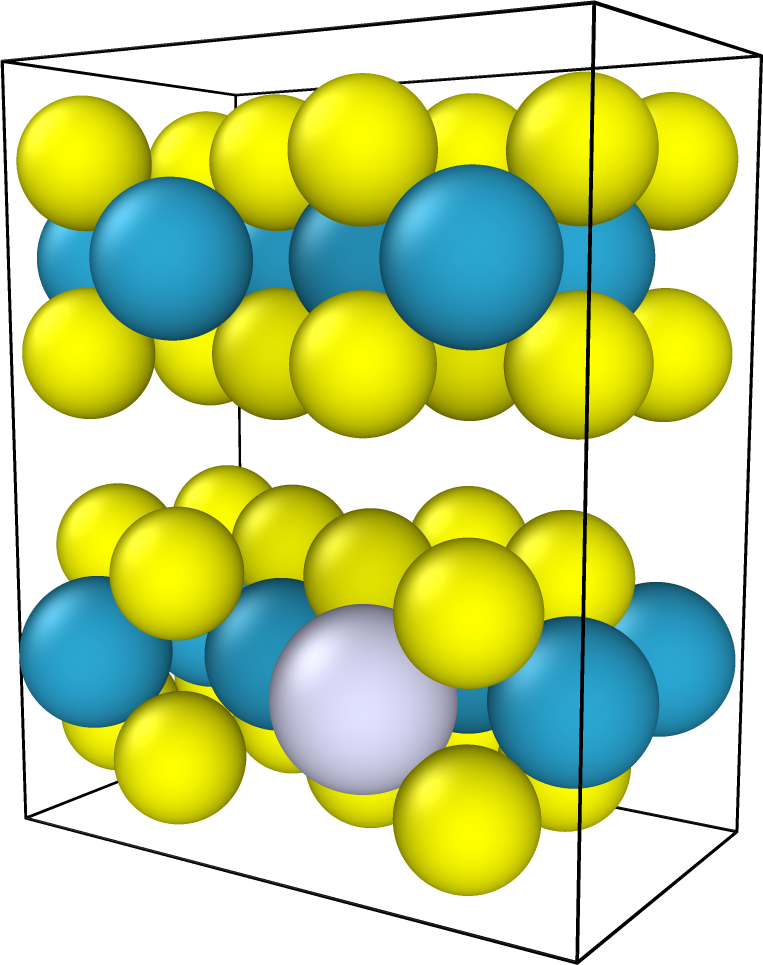}%
        \label{fig:mo_sub}%
    }
    \hfill
    \subfloat[Intercalated]{%
        \includegraphics[width=0.33\textwidth]{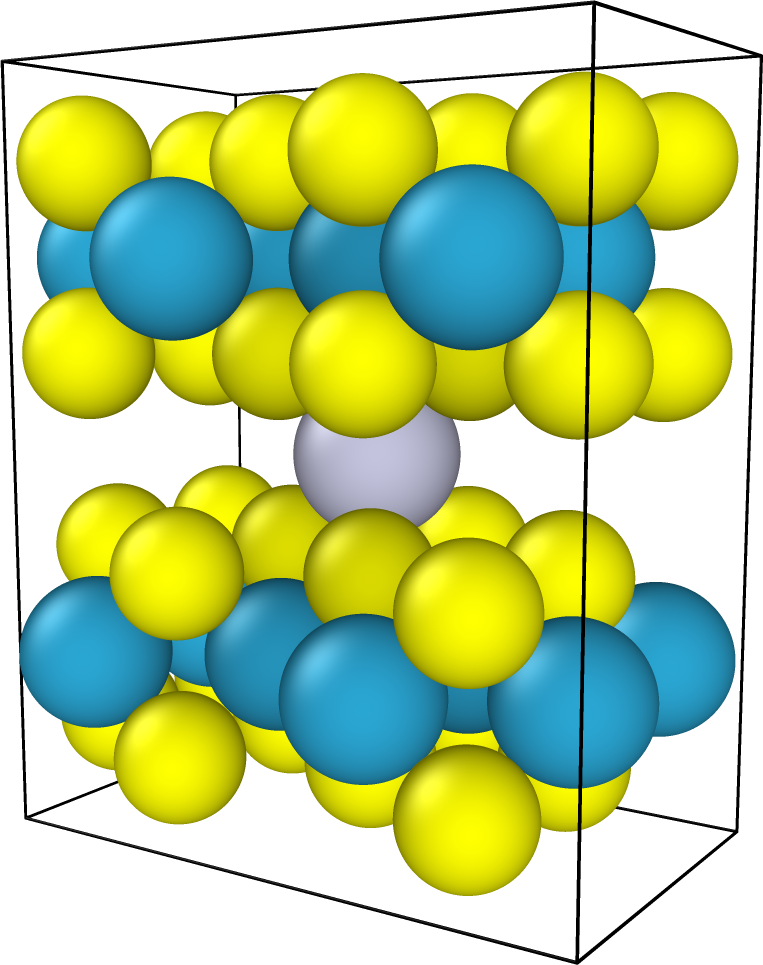}%
        \label{fig:intercalated}%
    }
    \caption{Snapshots of the three 48-atom \ce{MoS$_2$} prototypes: (a) S-site substitution, (b) Mo-site substitution, and (c) intercalation between layers. Mo is in blue, S in yellow, and the dopant in gray.}
    \label{fig:dopant_sites}
\end{figure*}

The dopant set is \ce{Ag}, \ce{Al}, \ce{Au}, \ce{C}, \ce{Cl}, \ce{Cu}, \ce{F}, \ce{Fe}, \ce{Ir}, \ce{Li}, \ce{N}, \ce{Na}, \ce{Nb}, \ce{O}, \ce{Pd}, \ce{Pt}, \ce{Re}, \ce{Rh}, \ce{Ru}, \ce{Si}, \ce{Ta}, \ce{Te}, \ce{Ti}, \ce{V}, and \ce{Zn}.
For each dopant \(X\), we built three 48-atom \ce{MoS$_2$} test structure: S-site substitution, Mo-site substitution, and intercalation between layers, as shown in Figure~\ref{fig:dopant_sites}. Each structure was relaxed separately with MLMD and with DFT.

\FloatBarrier

\label{sec:formation energy}
We compute the neutral formation energies $E_{\mathrm{form}}$ with the Zhang–Northrup formalism~\cite{Zhang1991-on}. For S-site substitution:
\begin{equation}
E^{X@\mathrm{S}}_{\mathrm{form}} 
= E_{\mathrm{tot}}(\mathrm{MoS_2}\!:\mathrm{S}\!\rightarrow\!X) 
- E_{\mathrm{tot}}(\mathrm{MoS_2}) + \mu_{\mathrm{S}} - \mu_{X}.
\end{equation}
For Mo-site substitution:
\begin{equation}
E^{X@\mathrm{Mo}}_{\mathrm{form}} 
= E_{\mathrm{tot}}(\mathrm{MoS_2}\!:\mathrm{Mo}\!\rightarrow\!X) 
- E_{\mathrm{tot}}(\mathrm{MoS_2}) + \mu_{\mathrm{Mo}} - \mu_{X}.
\end{equation}
For intercalation:
\begin{equation}
E^{X,int}_{\mathrm{form}} 
= E_{\mathrm{tot}}(\mathrm{MoS_2}\!+\!X) 
- E_{\mathrm{tot}}(\mathrm{MoS_2}) - \mu_{X}.
\end{equation}
Where $E_{tot}$ is total energy, $\mu$ is the chemical potential and $X$ indicates the dopant.
Within each method (MLMD or DFT), we use consistent elemental references. We set $\mu_{\mathrm{S}}=\tfrac{1}{8}E(\mathrm{S}_8)$ in a large box. We set $\mu_{\mathrm{Mo}}$ and $\mu_X$ from the lowest-energy elemental phase available in that method (e.g., bcc/fcc/hcp bulk, molecular box), selected per element based their lowest energy phase identified with Material Projects \cite{Jain2013-gm}. This one-method/one-reference scheme allows direct MLMD–DFT comparison of $E_{\mathrm{form}}$ values.
The code used for validation is available on github \cite{FaiyadUnknown-dv}.

It is important to note that the structure relaxation and formation energy calculation for 48-atom systems for all 25 dopants took between 0.53 to 0.97 hours of compute time using UMA small and 0.6 to 1.2 hours of compute time for UMA medium, depending on whether the calculations are run on the A100 (slower) GPU or L40 (faster) GPU. In contrast, the same calculation took about 410 hours of compute time with DFT (run on 56 CPU cores). So the MLIPs were about 341 to 820 times faster.

\subsection{MLIP simulations} 
\begin{figure}[ht]
    \centering
    \includegraphics[width=0.7\textwidth]{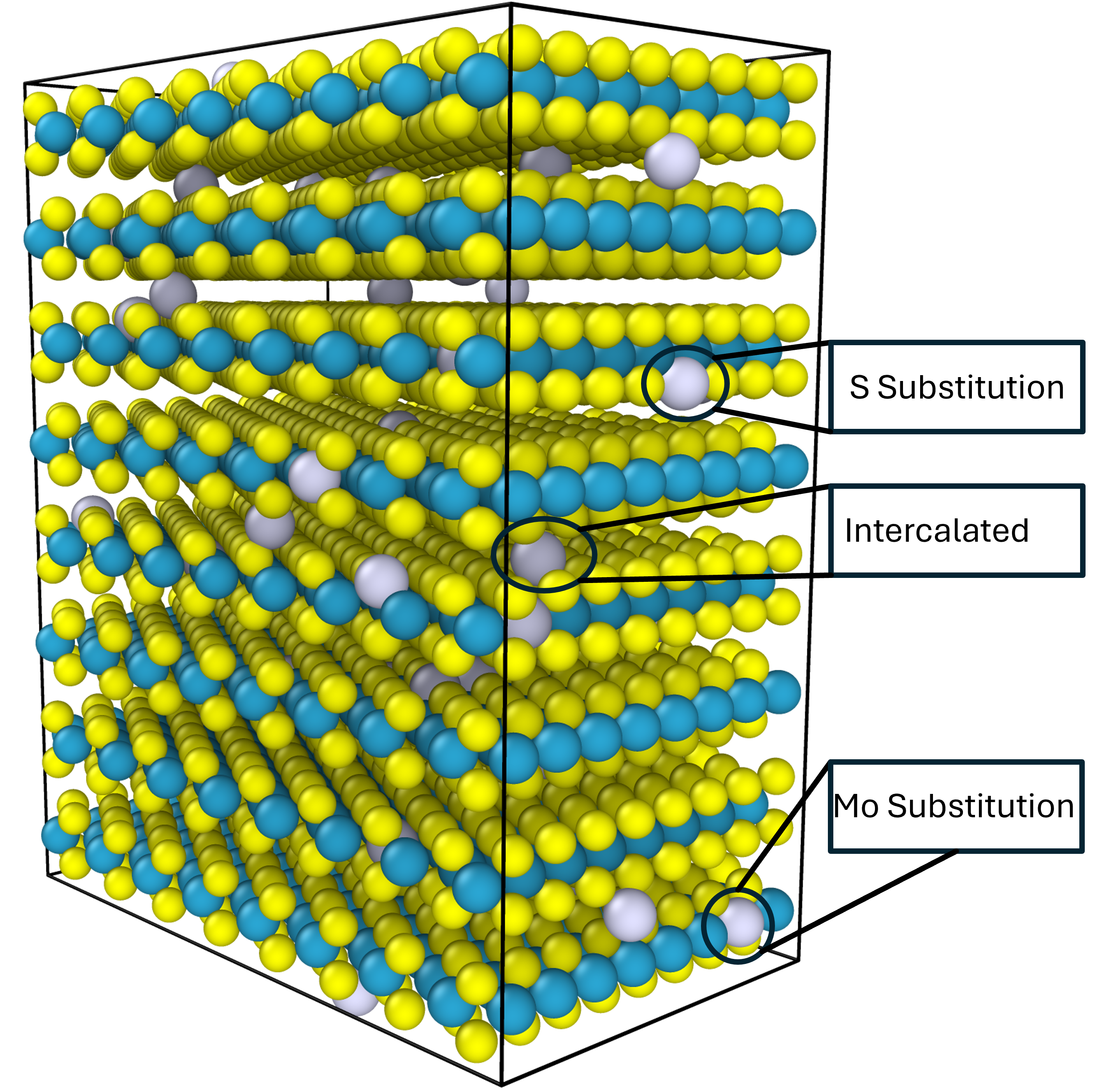}
    \caption{Snapshot of the bulk system used for MD simulation. The three initial sites of the dopant atoms in the MoS$_2$ are identified. Sphere colors are the same as in Figure~\ref{fig:dopant_sites}.}
    \label{fig:MD_System}
\end{figure}

The bulk MD system consists of an $8 \times 8 \times 4$ supercell containing approximately 3100 atoms and 8 layers of MoS$_2$. Dopants are introduced at an overall concentration of 5~wt\%, distributed approximately equally across three distinct doping sites: Mo substitution, S substitution, and intercalated (Figure~\ref{fig:MD_System}). For all simulations, the timestep is set to 1~fs. For NVT, a time constant for Berendsen temperature coupling (taut) value of 100~fs is used. 
An inhomogeneous NPT Berendsen ensemble with masking is used to allow anisotropic pressure equilibration at a target pressure of 1 atm along all axes. The temperature coupling time constant (taut) is set to 100 fs, and the Berendsen pressure coupling time constant (taup) is set to 500 fs.
%Inhomogeneous NPT Berendsen with masking to allow anisotropic pressure equilibration is used with taut of 100~fs and, a time constant for Berendsen pressure coupling (taup) value of 500~fs and a pressure of 1~atm. 

\FloatBarrier

Initial structures are optimized using the BFGS algorithm to minimize residual forces and stresses. Subsequently, the system is equilibrated in a series of steps to ensure thermodynamic stability and structural relaxation. The equilibration protocol involved: (i) an NVT ensemble using the Berendsen thermostat at 300~K, (ii) an NPT ensemble with anisotropic Berendsen barostat and thermostat at 300~K, and (iii) a final NVT ensemble at 300~K. Each equilibration phase is continued until convergence criteria are met, defined by temperature fluctuations less than 5~K over a 500~fs window. For NPT runs, an additional criterion required density fluctuations to be below 0.1~g/cm$^{3}$ over the same period.

%heating cycle
Following initial equilibration, the doped system is heated to 1000~K in an isothermal-isobaric (NPT) ensemble. Simulations are performed at 1000~K since that is below the temperature at which MoS$_2$ is reported to decompose \cite{Brainard1969-ch, Chen2019-xu}. Before heating, atomic velocities are initialized from a Maxwell-Boltzmann distribution at 300~K and the total linear momentum is removed prior to the start of the simulation. During heating, temperature is ramped linearly from 300~K to 1000~K over 20~ps (heating rate of 35~K ps$^{-1}$). An inhomogeneous NPT Berendsen barostat is used with an external pressure of 1~atm. Thermostat and barostat time constant of 100 fs and 500 fs, respectively, are used, and compressibility for the barostat is calculated based on the bulk modulus of MoS$_2$ of 80~GPa. The target temperature is updated every 10 integration steps to enforce a continuous thermal ramp.

After the heating cycle, the system is re-equilibrated at 1000~K for an additional 10~ps using the same thermostat-barostat parameters to relax the cell volume and density. Following this, a longer equilibration is performed for 100~ps in the canonical (NVT) ensemble at 1000 K using a Berendsen thermostat with a relaxation time of 100 fs. During this stage, the atomic mobility and dopent diffusion is analyzed. Equilibrium at each stage is assessed by monitoring the convergence and stability of the instantaneous temperature, total energy, and simulation cell volume.

After the high-temperature equilibration, the system is cooled from 1000 K to 300 K over 20 ps, corresponding to a cooling rate of -35 K ps$^{-1}$, in the NPT ensemble using the same pressure, thermostat, and barostat parameters used during heating. The temperature is again updated incrementally every 10 steps. The system is then equilibrated at 300 K in the NPT ensemble for 20 ps. Finally, a 100 ps NVT simulation at 300 K is performed.
%to qualitatively assess the relaxed local structure and dopant configurations. The average density is calculated from the final 10 ps of this trajectory.
%Following equilibration, a heating cycle is performed. The system is heated from 300~K to 1000~K over 20~ps followed by equilibration at 1000~K for 20~ps in an NPT ensemble. Next the system is equilibrated at 1000~K for 100~ps in an NVT ensemble. while providing sufficient thermal energy for the dopant to diffuse and the local structure of MoS$_2$ to respond to the mobility of the dopant.
%During these simulations, the mean square displacement (MSD) is calculated (details in SI Sect. S1) to quantify atomic mobility and dopant diffusion.
Trajectory data from this last stage are processed using custom Python scripts based on the Atomic Simulation Environment (ASE).
The MLIPs enabled MD simulations of $\sim$3100 atoms. The same simulation cannot be performed with DFT using typical computing resources.
%cooling cycle
%After the heating cycle, a cooling cycle is performed on the system. During this cycle, the system is cooled from 1000~K to 300~K over 20~ps followed by equilibration at 300~K for 20~ps in an NPT ensemble. From the last 10~ps of this NPT cycle, we calculate the average density of each doped MoS$_2$ system.
% 300 K NVT and mobility calculation
%Finally, the system is re-equilibrated at 300~K for 100~ps in a NVT ensemble. 
%These trajectories are analyzed qualitatively to understand how different dopants behave in and affect the MoS$_2$ nanostructure.
All scripts used for performing these simulations are available on github. \cite{FaiyadUnknown-dv}

\section{\label{sec:results}Results and Discussion}

\subsection{\label{sec:results:subsec:Validation}Validation of UMA potentials}

\begin{figure*}[ht]
    \centering
    \subfloat[UMA Small parity plot]{%
        \includegraphics[width=0.48\textwidth]{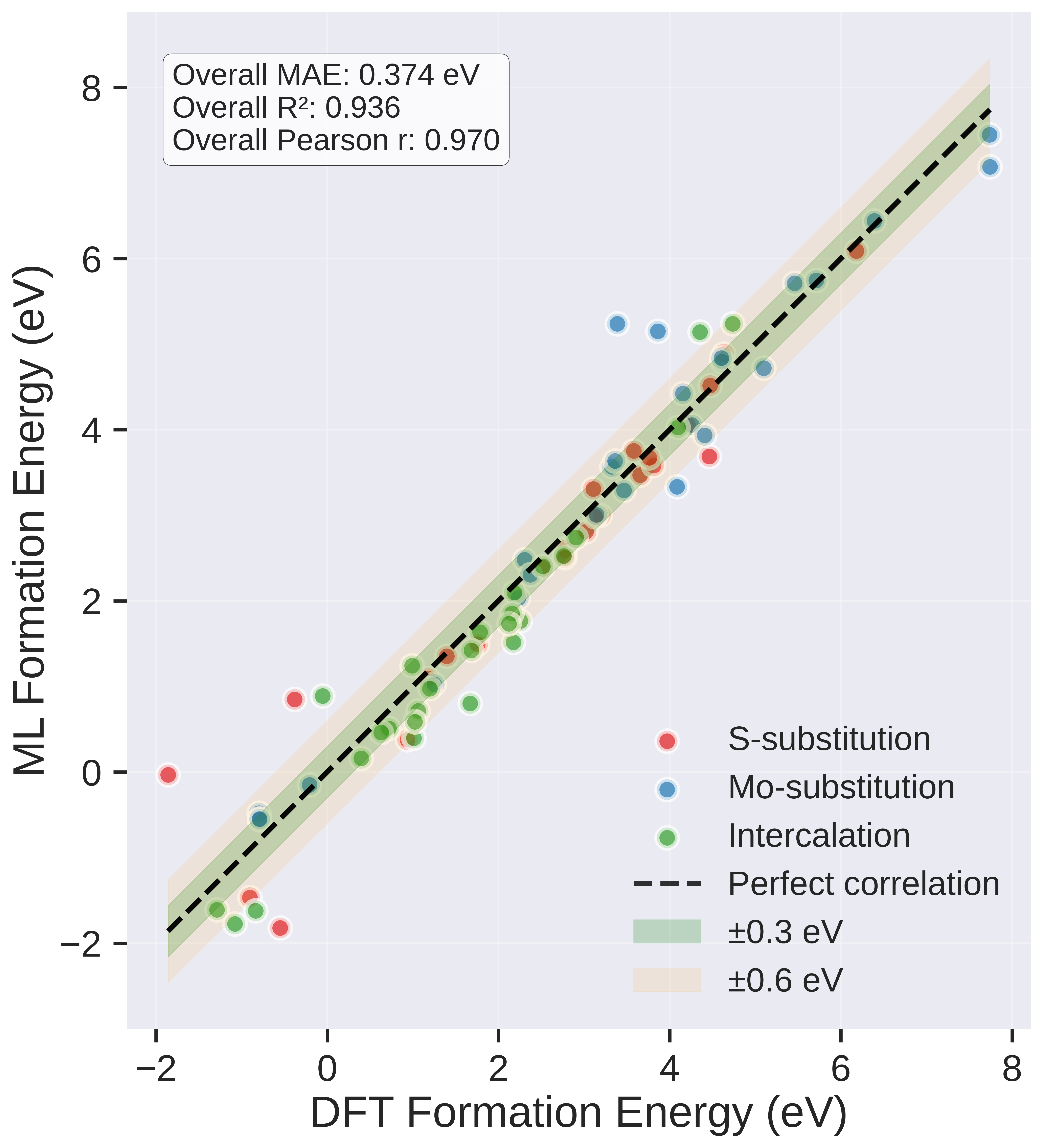}%
        \label{fig:uma_sm_validation_comb_parity_plot}%
    }
    \hfill
    \subfloat[UMA Medium parity plot]{%
        \includegraphics[width=0.48\textwidth]{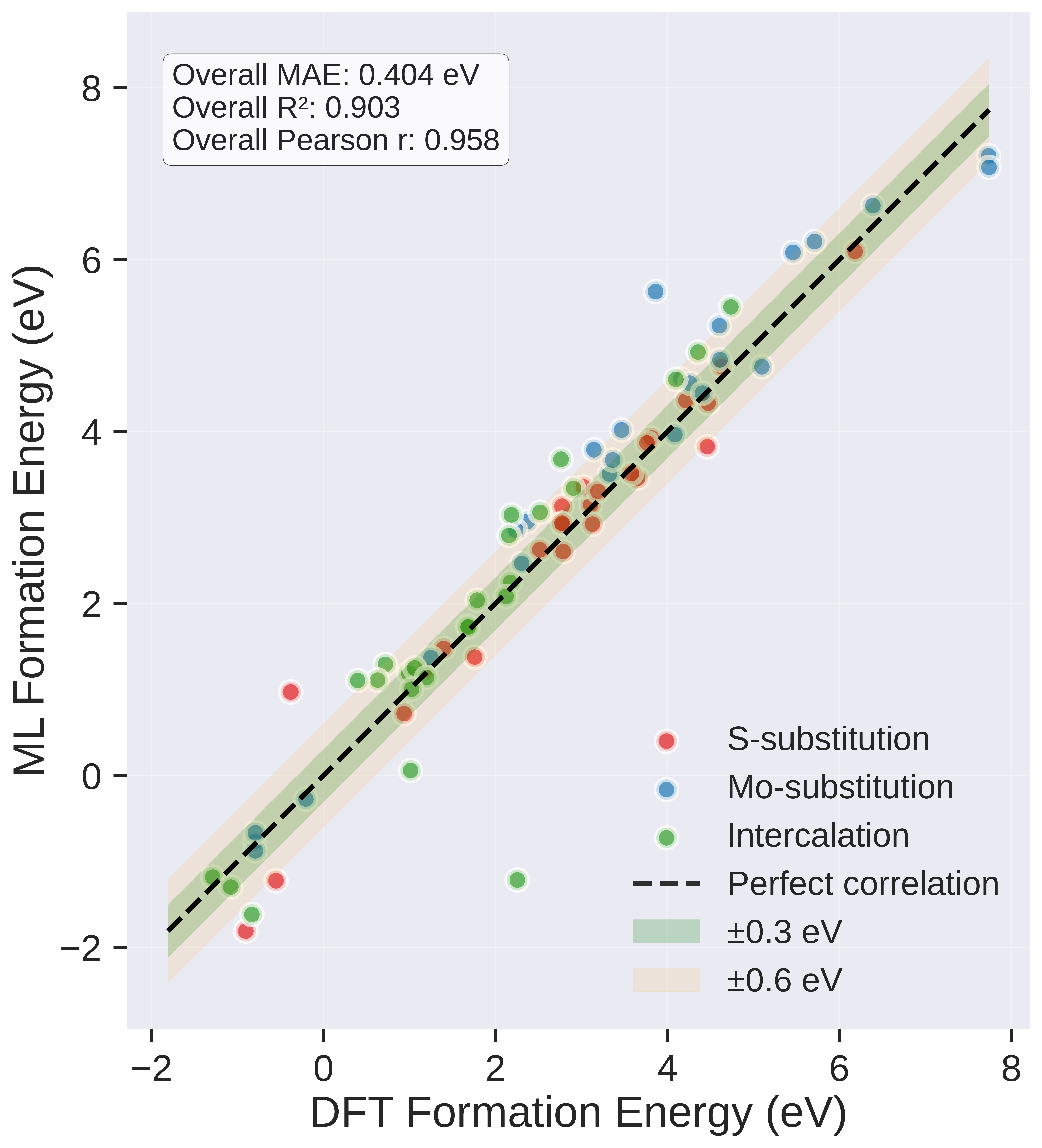}%
        \label{fig:uma_m_validation_comb_parity_plot}%
    }
    \caption{Parity plots comparing formation energies calculated by (a) UMA small and (b) UMA medium machine-learning interatomic potentials with DFT reference values for MoS$_2$ with 25 different dopants at three different positions in the MoS$_2$ lattice. UMA small has an overall MAE of 0.374~eV and UMA medium has an overall MAE of 0.404~eV.}
    \label{fig:Validation_parity}
\end{figure*}

% parity plots, MAE and R^2 
The accuracy of the UMA potentials is assessed to establish their reliability for modeling doped MoS$_2$. Figure~\ref{fig:Validation_parity} shows parity plots comparing formation energies computed by the UMA small and UMA medium models against reference DFT calculations based on the Zhang-Northrup formulation. The parity analysis shows that the mean average error (MAE) for the entire dataset is 0.374 \texttt{eV} for UMA small and 0.404 \texttt{eV} for UMA medium . For UMA small, the MAE is 0.377~eV for the S-substitution, 0.360 \texttt{eV} for the Mo-Substitution, and 0.326~eV for the intercalated case. For UMA medium, these values are 0.277 \texttt{eV} for S-substitution, 0.398 \texttt{eV} for Mo-substitution, and 0.536 \texttt{eV} for intercalated case. In both UMA small and UMA medium, the Pearson r values are >0.9, indicating strong positive linear relationship with DFT, and the R$^2$ values were >0.9, indicating that both models can accurately capture the energy change due to doping with different elements. 

\FloatBarrier

\begin{figure*}[ht]
    \centering
    \subfloat[Stacked error bars for UMA Small]{%
        \includegraphics[height=0.38\textheight]{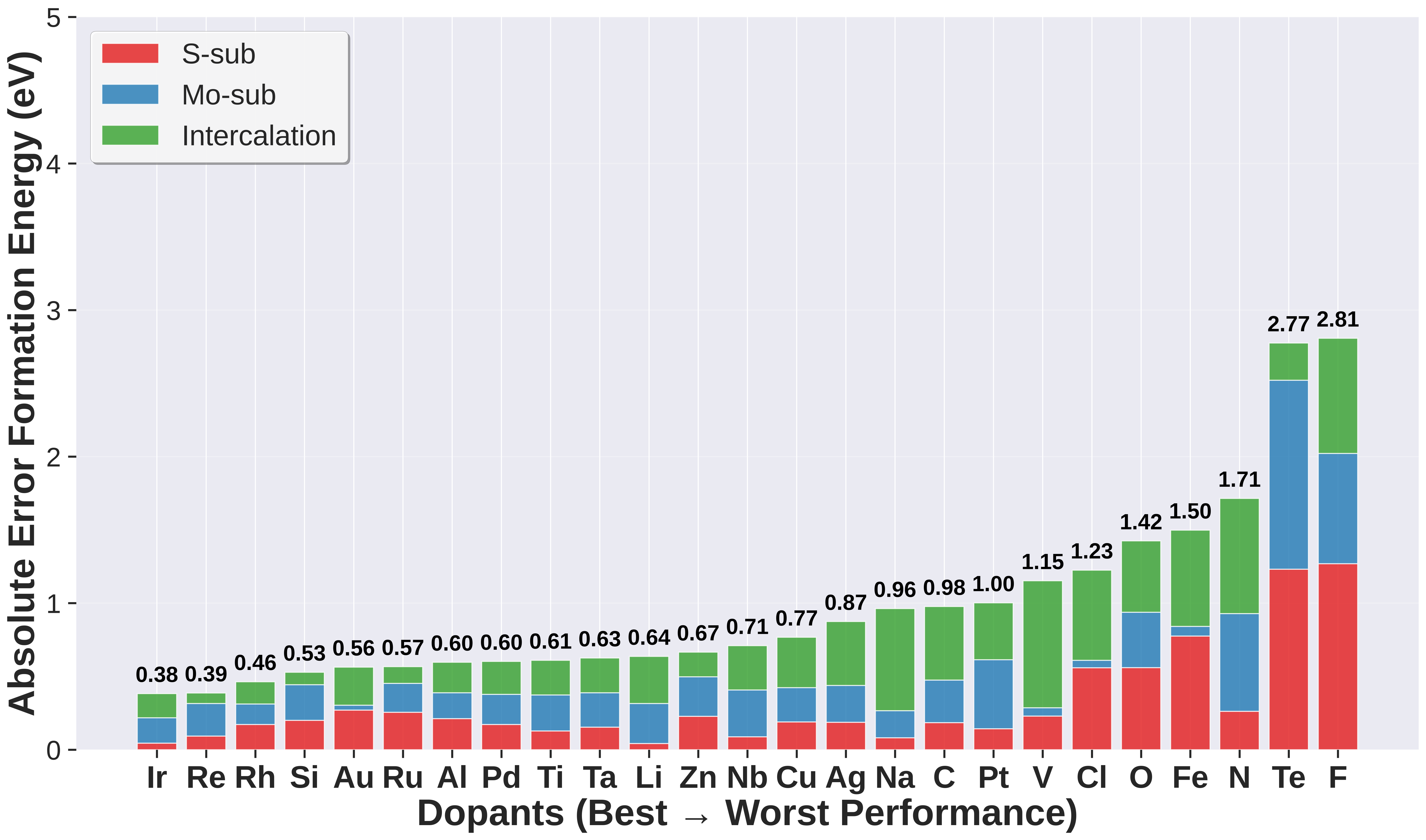}%
        \label{fig:stacked_error_bars_s}%
    }
    \par\medskip
    \subfloat[Stacked error bars for UMA Medium]{%
        \includegraphics[height=0.38\textheight]{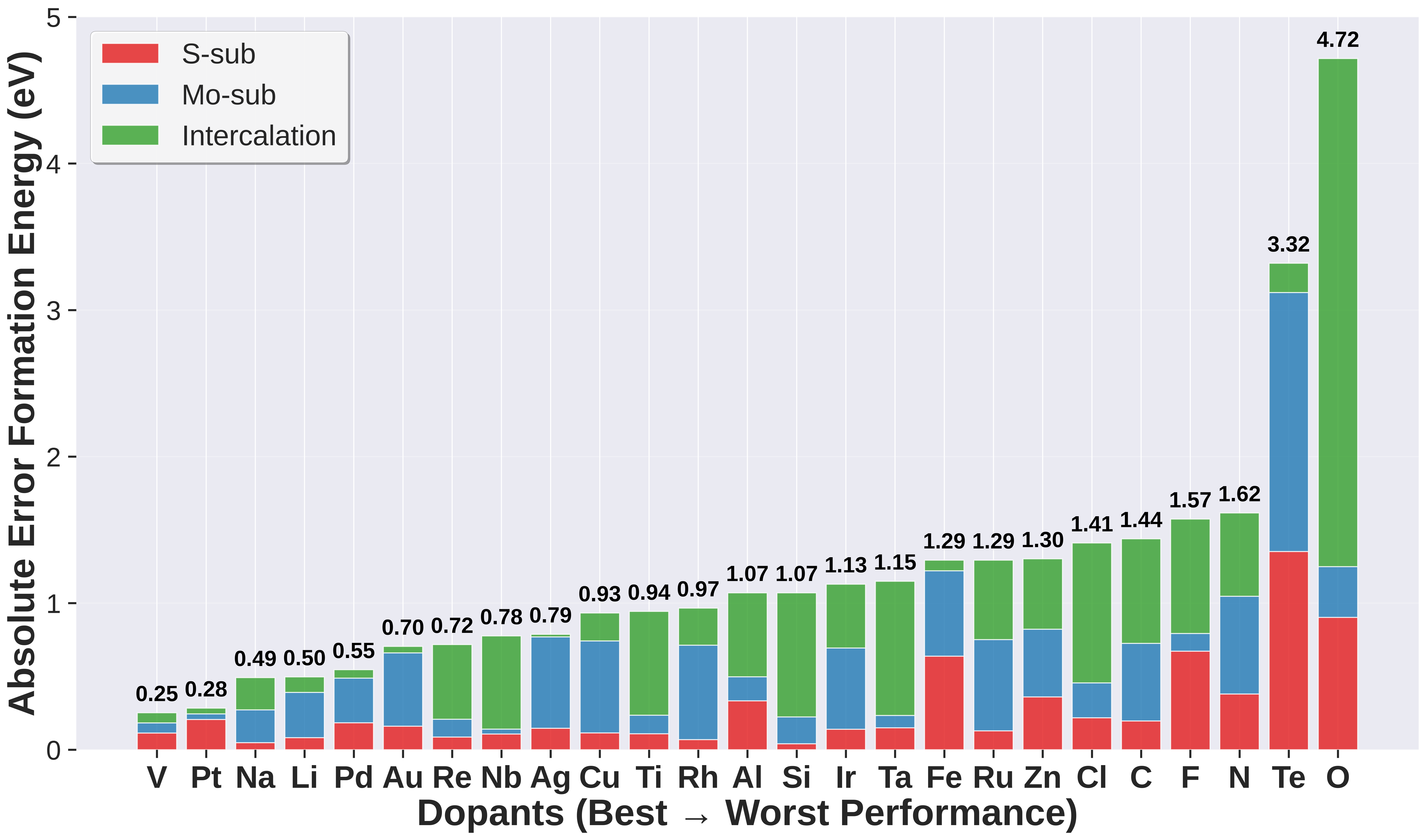}%
        \label{fig:stacked_error_bars_m}%
    }
    \caption{Stacked error bar plots of the magnitude of formation energy errors relative to DFT reference values at all three dopant sites for (a) UMA small and (b) UMA medium models. Dopants are plotted in order of increasing sum of absolute error and the three colors of each bar correspond to the three dopant sites in the MoS$_2$ structure.}
    \label{fig:stacked_error_bars}
\end{figure*}

% stacked error bar plot
The dopant-specific error magnitudes for both models are further detailed in the stacked bar plots of Figure~\ref{fig:stacked_error_bars}.
This analysis shows that the ML model achieves low error for many dopants. For UMA small, 19 of the 25 dopants tested have a cumulative absolute error of less than 1 \texttt{eV} and, for UMA medium, 11 of the 25 dopants tested have a cumulative absolute error of less than 1 \texttt{eV}. These errors are consistent with the defect formation energy errors observed for other universal MLIPs \cite{Berger2025-ua, Broberg2023-rm}. These results indicate that, for the test system used for validation, UMA small has a better overall accuracy compared to UMA medium. Additionally, for our test systems, UMA small is almost twice as fast as UMA medium in steps per second. Due to the better performance of UMA small, we use this MLIP for the remainder of the work. 

%\FloatBarrier

% analysis by metal class
Further analysis highlights trends across different dopant classes (Figure S1). 
Metal dopants consistently have the smallest deviation from DFT (MAE <0.3~eV in individual dopant locations), suggesting that the MLIP can simulate these metallic substitutions with good accuracy. In contrast, the non-metal dopants have larger errors (MAEs on the order of 0.5~eV across individual dopant sites), indicating that the MLIP generally provides a less accurate prediction of the formation energies of non-metal dopants.

% local structure analysis
We also examine the local structural accuracy of the UMA small potential by analyzing partial radial distribution functions (RDFs) for Mo-dopant distances for S substitution and S-dopant distances for Mo substitution cases. 
The position of the first peak in the RDF plots is used to approximate the nearest neighbor distance.
In the intercalated cases, the dopant atoms lie between MoS$_2$ layers so they are not included in this analysis. 
\begin{figure*}[ht]
    \centering
    \includegraphics[width=0.85\textwidth]{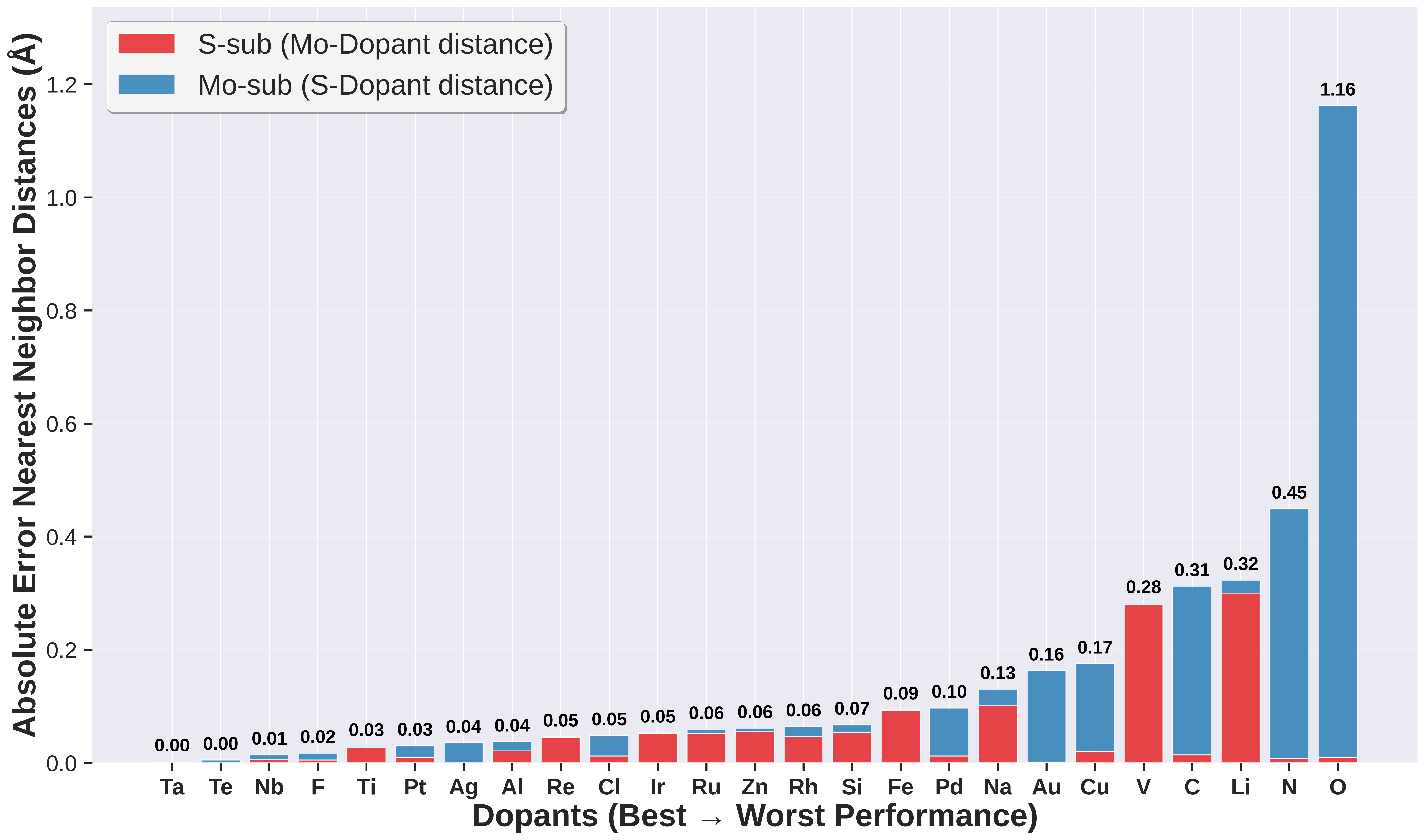}
    \caption{Absolute difference of distances predicted by UMA small and DFT between dopant and nearest neighbors for S substituted and Mo substituted cases, ordered from best to worst.}
    \label{fig:stacked_bond_length_errors}
\end{figure*}
%\FloatBarrier

The difference between the nearest neighbor distances from UMA small and DFT for each dopant is plotted in Figure~\ref{fig:stacked_bond_length_errors}. For all dopants up to Na, the error for any individual substituted system is <0.1~\AA\ (<3\% error). For Au and Cu (Mo substitution cases) this error increases to 0.16~\AA\ (6\% error). V, C, Li, N, and O show the highest error of 0.28~\AA\ to 1.15~\AA\ (10\% - 42\%) for individual cases.

From Figures~\ref{fig:stacked_error_bars} and \ref{fig:stacked_bond_length_errors}, we consider a few cases where the MLIP does not provide accurate energy and local structure predictions. These errors are likely caused by the fact that the UMA training dataset contains neutral bulk systems and does not explicitly include point defects, which could induce errors in energy and localized structure \cite{Barroso-Luque2024-xh}.
First, small, highly electronegative dopants at Mo sites, i.e., O and N substituting Mo, have the largest local structure error. DFT shows that these dopants create strong, localized bonds and significant lattice contraction. The ML model does not capture this extreme distortion or the associated energy change.
Second, alkali metals like Li or early transition metals like V substituting S led to large error, primarily in geometry. These dopants are much bigger than the S atom they replace, so DFT shows the local Mo–dopant bonds increasing in length. The MLIP partially failed to account for this expansion.
%, predicting bonds \~0.3 Å shorter than DFT (a \~10\% error). 
Interestingly, the formation energy error for these cases is not very large (<0.3~ \texttt{eV}) – meaning the model captures the thermodynamics of the system, even though it does not accurately predict the relaxed structure. 
Third, substituting S with Te is essentially alloying to form MoTe$_2$-like local environments. DFT shows this process is favorable (negative formation energy; Figure S2), whereas the MLIP severely underestimates that favorability (predicting positive formation energy; Figure S3).
Lastly, while dopants that are similar in character to the host elements (metals substituting Mo, or semi-metals substituting S) are predicted with good accuracy by the ML model, there are a few exceptions, for example, Fe at an S substitution site.
The error that does occur for metals tends to be smaller and possibly due to effects like magnetism or charge state differences which are not accounted for in the model training data. Thus, there are still opportunities to improve the accuracy of the model through finetuning of the MLIP. This will be explored in a subsequent study.

%Dopant Ranking
The results presented in Figures~\ref{fig:stacked_error_bars} and \ref{fig:stacked_bond_length_errors} demonstrate that UMA small predicts the formation energies of Ag, Al, Au, C, Cu, Ir, Li, Na, Nb, Pd, Pt, Re, Rh, Ru, Si, Ta, Ti, Zn, and Pt dopants in MoS$_2$ with formation energy errors below 0.3 eV for each of the three dopant sites. Furthermore, the structural effects of Ag, Al, Au, Cl, Cu, F, Fe, Ir, Na, Nb, Pd, Pt, Re, Rh, Ru, Si, Ta, Te, Ti, and Zn doping at the two substitution sites are reproduced with good accuracy, exhibiting deviations of less than 6\% relative to DFT. When both energetic and structural accuracies are jointly considered, the dopants Ag, Al, Au, Cu, Ir, Na, Nb, Pd, Pt, Re, Rh, Ru, Si, Ta, Ti, and Zn are identified as being most reliably captured by the UMA small model.

\subsection{\label{sec:results:subsec:Comparing Dopants} Demonstration of the MLIP}

\begin{figure*}[ht]
    \centering
    \includegraphics[width=0.85\textwidth]{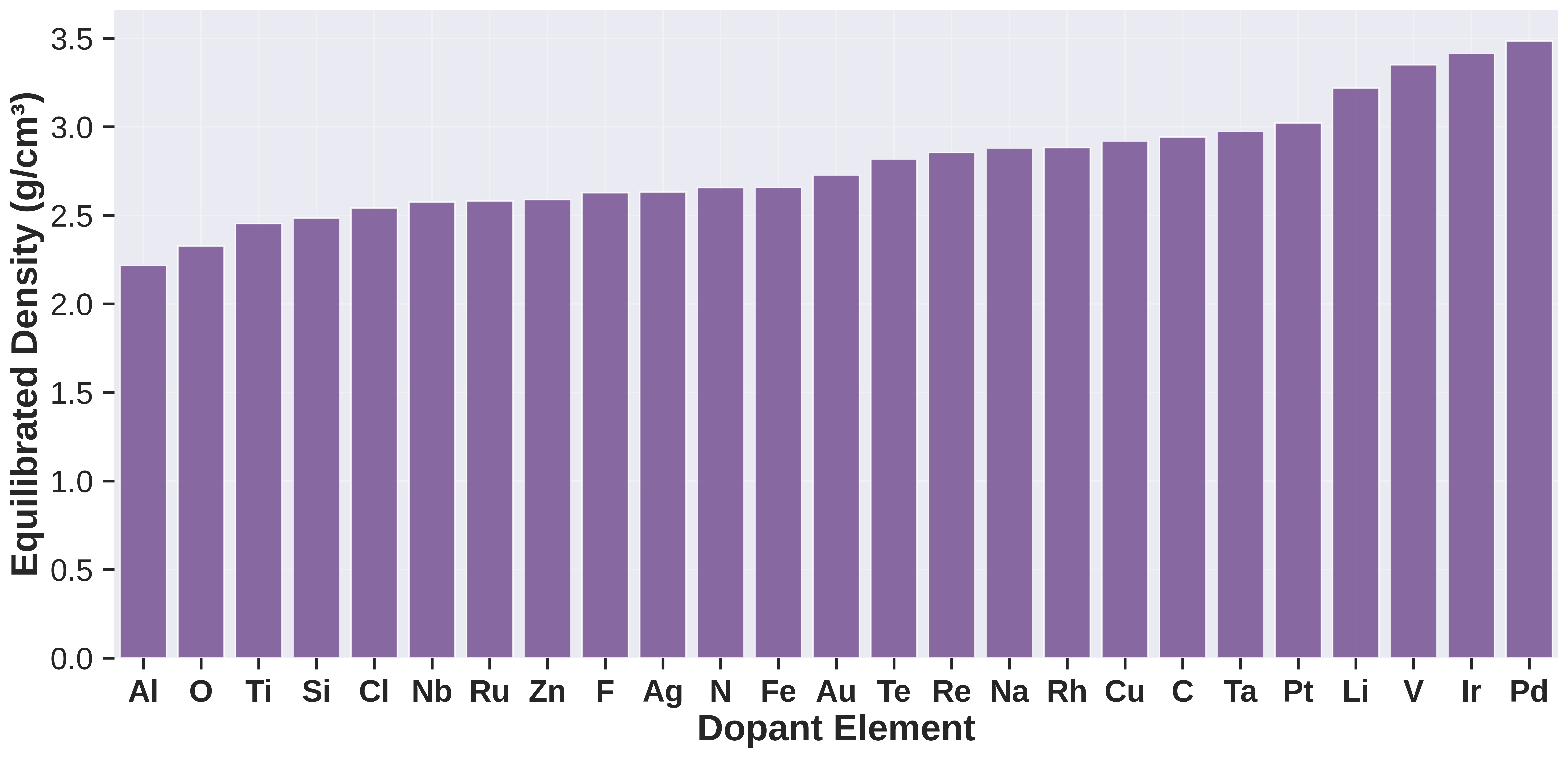}
    \caption{Doped MoS$_2$ densities calculated from the MLIP simulations at 300 K (after heating and cooling simulations), in order of increasing density.}
    \label{fig:density_distribution}
\end{figure*}

The density of the doped systems calculated from the MLIP simulations at 300 K are shown in Figure~\ref{fig:density_distribution}.   The densities of the simulated doped MoS$_2$ systems are between 2.2 and 3.5~g/cm$^3$.
There is a wide range of densities for MoS$_2$ reported in literature. For sputter deposited coatings, a density range from 1.90 to 5.29~g/cm$^3$ has been reported \cite{Babuska2025-dk}. 
Regardless, the fact that the model densities are reasonable compared to the large experimental range is encouraging for this demonstration of the MLIP simulations.

\FloatBarrier

%\subsection{Dopant Mobility from Mean Square Displacement Analysis}
The diffusivity of the dopants is quantified by the slope of the MSD vs. time data at 1000~K. This parameter is a measure of the stability of the dopant in the MoS$_2$ lattice and an indicator of dopant migration and clustering. Lower diffusivity suggests that the dopant is likely to remain at its initial site, while higher diffusivity indicates dopants are mobile, which can affect the nanostructure of the material.

\begin{figure}[ht]
    \centering
    \includegraphics[width=0.85\textwidth]{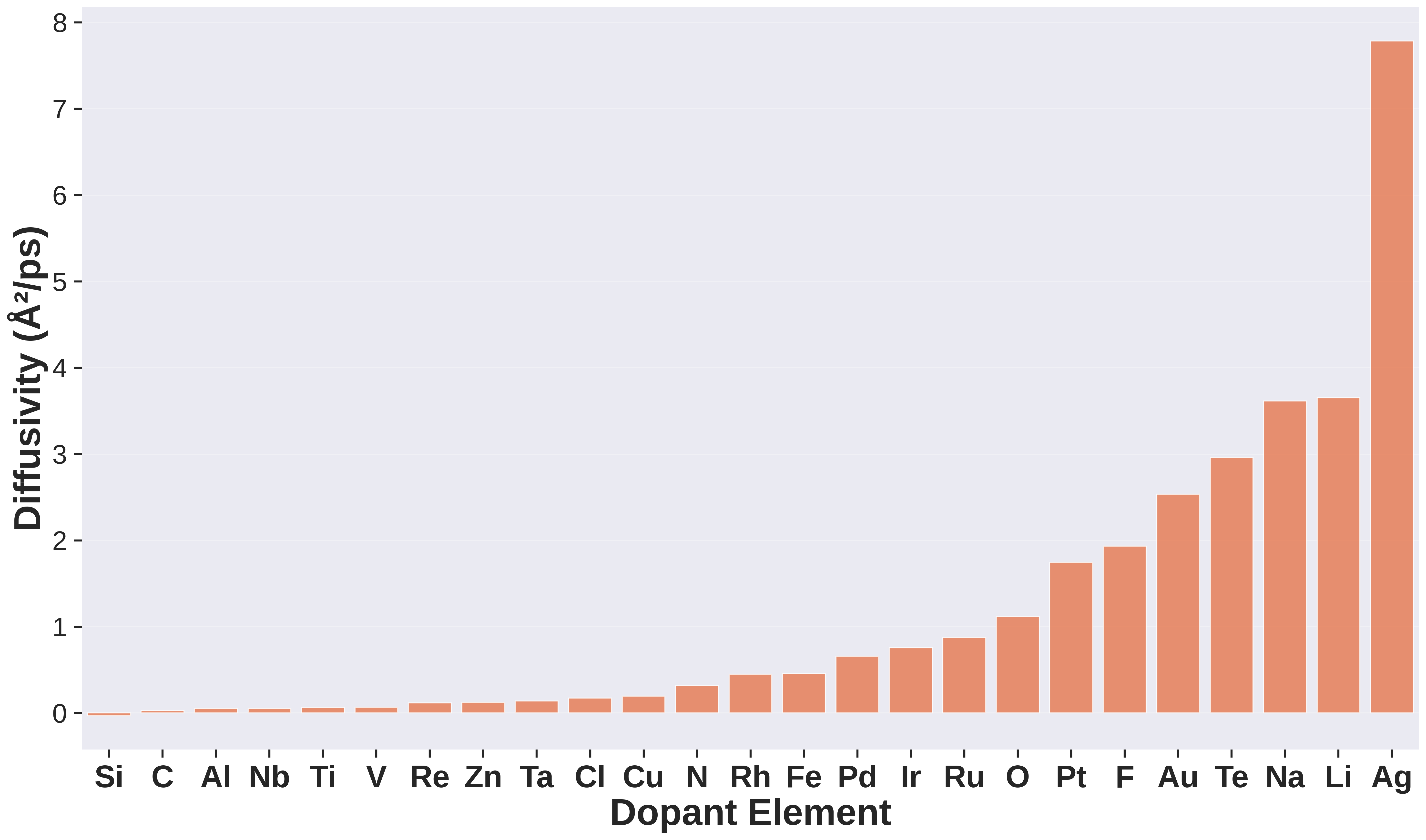}
    \caption{Diffusivity for each dopant element determined by linear fitting of MSD vs. time data from the last half of the MLIP simulation at 1000~K, in order of increasing diffusivity.}
    \label{fig:msd_slopes}
\end{figure}

Figure~\ref{fig:msd_slopes} summarizes the diffusivity for all examined dopants, providing a quantitative comparison of relative mobilities.
The largest diffusivity is exhibited by Ag ($\sim 7.8~\textrm{\AA}^2/\textrm{ps}$), followed by Li and Na ($\sim 3.6~\textrm{\AA}^2/\textrm{ps}$), indicating that these dopants are highly mobile at 1000~K.  Moderate diffusion ($1$--$2~\textrm{\AA}^2/\textrm{ps}$) is observed for dopants from O to Te in Figure \ref{fig:msd_slopes}. However, most dopants, Si to Ru in Figure \ref{fig:msd_slopes}, exhibit nearly negligible diffusivity $\sim 1~\mathrm{\AA^2\,ps^{-1}}$, reflecting limited kinetic motion during the simulation.

\FloatBarrier

We classify the dopants into four groups based on their diffusivity, MSD and RDF plots (Supporting Information S4 - S7), and visual observation of dopant behavior during the simulation at 1000~K. These groups are: metals that form clusters, metals that do not cluster, light metals that diffuse through MoS$_2$, and non-metals that chemically interact with MoS$_2$. 
We choose one representative dopant to analyze in detail and illustrate the behavior characteristic of each group.
Note that the dopants presented in the subsequent qualitative analysis are chosen not necessarily because they have the least formation energy or atom distance error, but because they most clearly exhibit behavior characteristic of a given dopant group.

%Metals that form clusters
The first group of dopants shows clustering behavior, where initially distributed dopant atoms exhibit a strong tendency to aggregate during the simulation. 
Clusters are identified visually as groups of three or more dopant atoms.
Of the dopants simulated, Al, Cu, Fe, Ir, Nb, Pt, Re, Rh, Ru, Ti, Ta, V, and Zn exhibit this clustering behavior. 
Larger clusters are formed for dopants with lower atomic weight, which can be attributed to the greater number of dopant atoms with lower atomic weight (since dopants constitute 5 wt\% of the system).
All dopants in this group have very low mobility (<1 Å$^2$/ps).
To understand the behavior of the dopants in this group, we analyzed Cu-doped MoS$_2$.

%Ranking
\begin{figure}[ht]
    \centering
    \includegraphics[width=\textwidth]{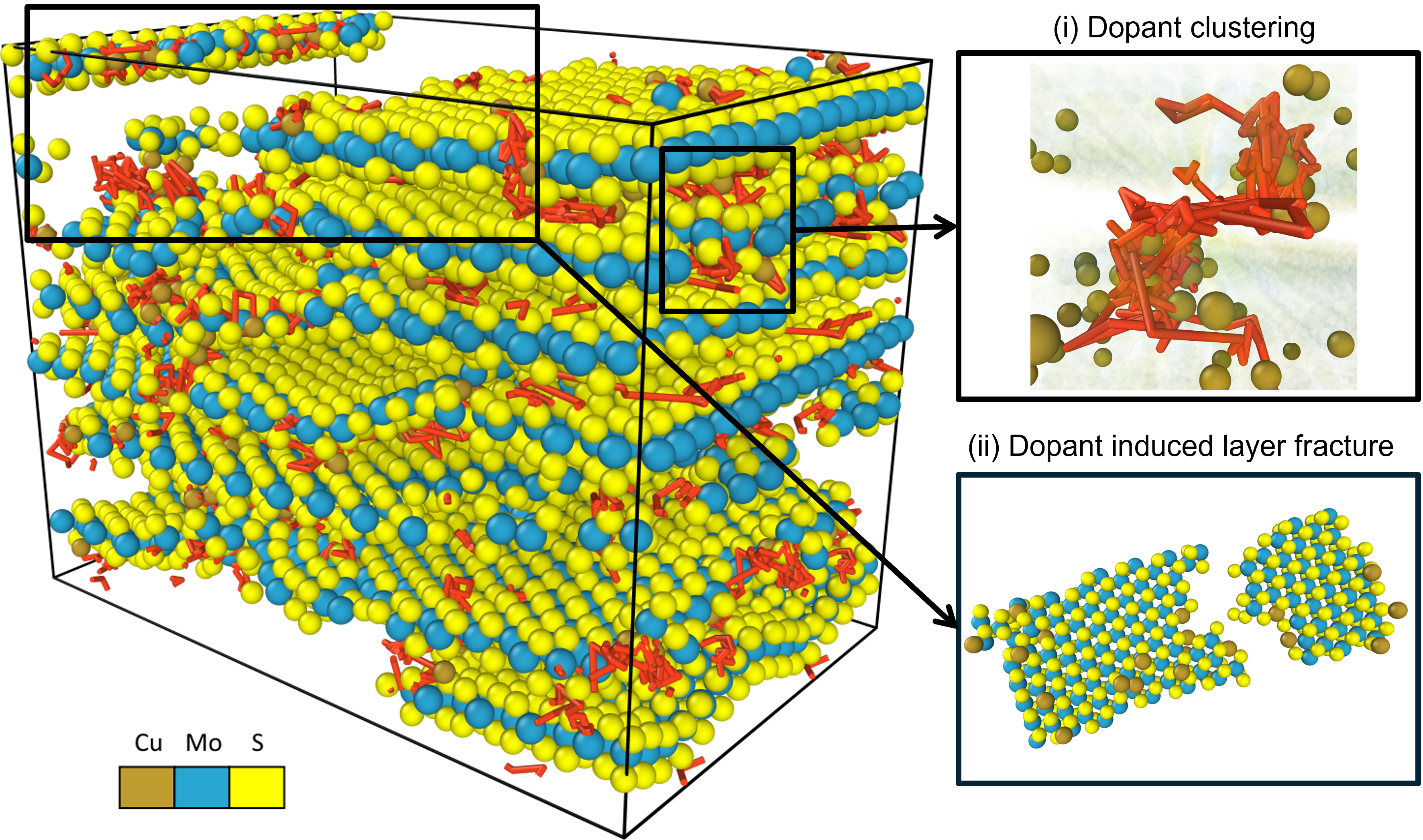}
    \caption{Snapshot from the Cu-doped MoS$_2$ MLIP simulation highlighting behavior that is representative of metal dopants that form clusters. In the main figure and inset (i), red lines indicate 100 ps dopant trajectories that show the process of cluster formation. In (ii), a single layer of MoS$_2$ from the bulk simulation is shown to highlight the fracture of MoS$_2$ layers for some doped systems.}
    \label{fig:Cu_doped_MoS2}
\end{figure}
\FloatBarrier

In Figure \ref{fig:Cu_doped_MoS2}, a snapshot of the Cu-doped system is shown with red lines indicating the dopant trajectories during the 100~ps NVT equilibration at 1000~K. In this system, two types of dopant behavior are observed, depending on the initial position in the MoS$_2$ lattice. First, substitutional dopants are very stable and rarely diffuse away from their initial positions. 
In contrast, intercalated dopants have higher mobility, moving freely throughout the lattice via thermal motion. This diffusion is the primary mechanism driving cluster formation. 
Cluster formation generally starts when an intercalated dopant nears a substitutional dopant. 
Figure \ref{fig:Cu_doped_MoS2}(i) illustrates this behavior with a close up snapshot of a cluster of Cu dopant atoms formed during the simulation. Red trajectory lines show the intercalated dopants moving to form a cluster. 
Clustering of dopants in MoS$_2$ has been reported previously in the literature. For example, Re dopants in MoS$_2$ exhibit significant clustering at dopant concentrations of approximately 2 atom\%, forming aggregates into $\sim$3 nm phase-segregated domains at 10 atom\% doping, as revealed by Z-contrast scanning transmission electron microscopy (Z-STEM) and scanning tunneling microscopy (STM) analysis \cite{Munson2025-vs}.
In our simulations, once clusters are formed, the dopants become immobile, which leads to the all elements in this dopant group having low diffusivity (Figure \ref{fig:msd_slopes}).

Beyond cluster formation, some dopants in this group induced fracture in the MoS$_2$ layered structure, as illustrated in Figure~\ref{fig:Cu_doped_MoS2}(ii). This snapshot shows representative behavior where there are many Cu atoms near the fractured edge. For undoped MoS$_2$, fracture is not observed (Figure S8). 
These observations suggest that dopant-host interactions can cause fracture in the layers, which can compromise the structural integrity of the MoS$_2$ layers. 
This is consistent with previous experiments in which cracking and delamination were more pronounced in Ni-doped MoS$_2$ than undoped MoS$_2$ during tribotesting \cite{Vellore2020-hv}. 

%Metals that do not form clusters
The second group of dopants is metals that do not exhibit clustering at 1000~K. This behavior is observed in the MLIP simulations with Ag, Au, and Pd dopants. Unlike the previous group, the intercalated dopants of this group do not cluster together when in close proximity to other intercalated or substitutional dopants. 
This lack of clustering means that the intercalated dopants remain mobile throughout the simulation. As a result, these dopants have higher diffusivity than the clustering metals group (2 to 9 times higher). 
The behavior observed in these simulations is consistent with previous reports that Pd dopants form clusters at lower temperatures but, at temperatures between 700~K to 1200~K, the clusters disappear and disperse within the lattice \cite{Kamaratos1987-yh}.
No fracture of the MoS$_2$ is observed in the simulations with this group of dopants. The lack of fracturing may be a factor contributing to the observation that Au-doping increases MoS$_2$ crystallinity in a previous experimental study \cite{Scharf2013-lz}.

%Light metals that have interlayer diffusion
\begin{figure}[ht]
    \centering
    \includegraphics[width=\textwidth]{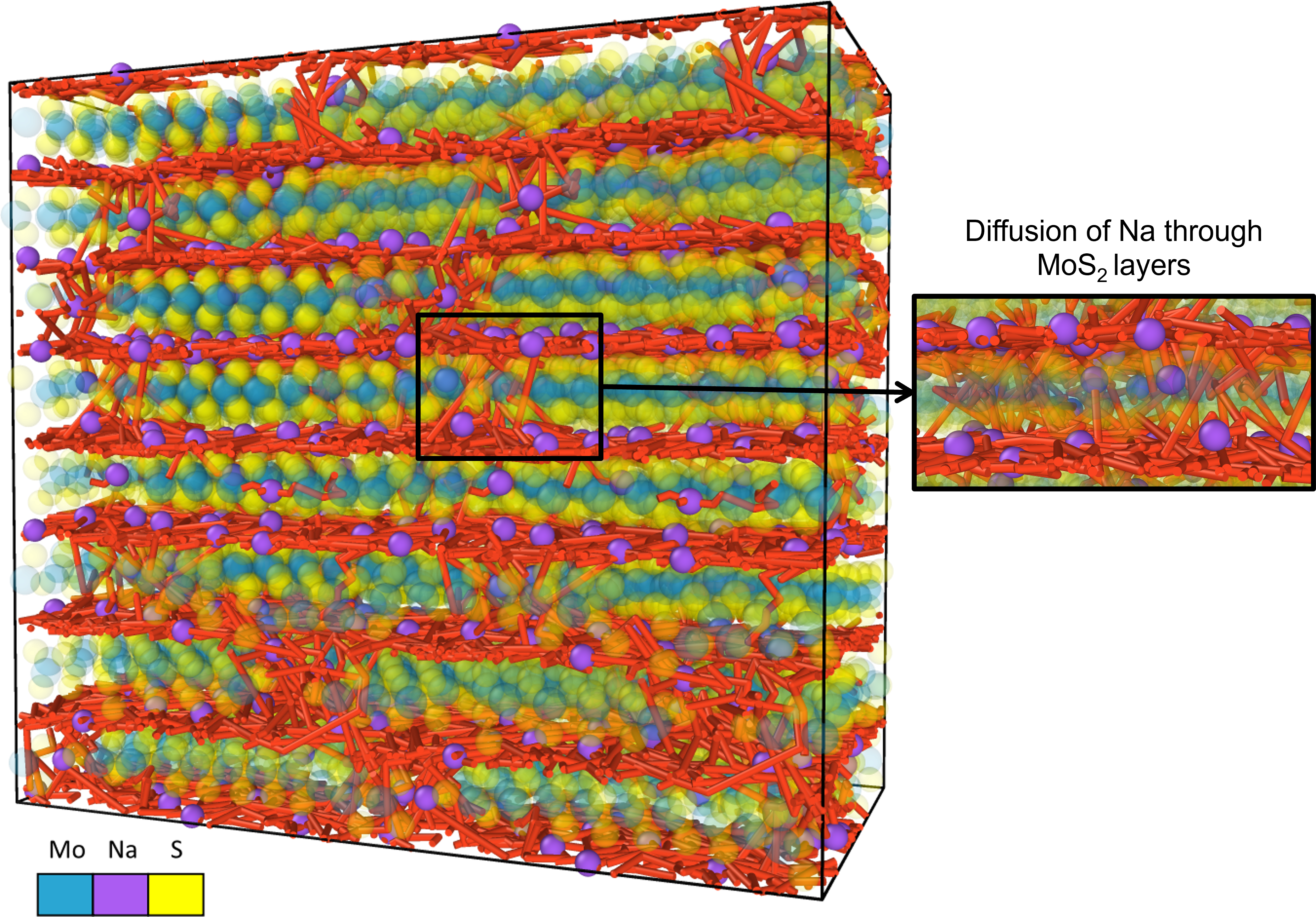}
    \caption{Snapshots from the Na-doped MoS$_2$ MLIP simulation with Mo and S made transparent to highlight the interlayer diffusion of the light metal dopants. The red lines indicate the dopant trajectories over 100~ps. The closeup view highlights diffusion of the Na through the MoS$_2$ layers.}
    \label{fig:diffusion bridge}
\end{figure}
\FloatBarrier

The third group of dopants comprises two light metals, Li and Na. These dopants do not show the substitutional stability of the elements in the previous two groups. Also, similar to the second group, this group lacked any clustering behavior at 1000~K. Instead, Li and Na exhibit significant diffusion, both in the intercalated space between layers as well as through the MoS$_2$ layers, as shown in Figure \ref{fig:diffusion bridge}. The small atomic radii and light atomic weight of these two dopants facilitates diffusion. 
The intercalated diffusion behavior that we observe is consistent with the previously reported low energy barrier for Na diffusion between intercalated sites in both pristine MoS$_2$ and MoS$_2$ heterostructures (0.16 eV to 0.53 eV depending on diffusion direction) \cite{Yao2019-sd, Chen2024-vr}.
Our simulations also show that the substitutional dopants diffuse out from their initial positions, leaving behind a vacancy. Then, intercalated dopants from the adjacent layers fill those vacancies and are free to diffuse out of the substitutional position again. 
This interlayer diffusion pathway can be explained based on previous DFT studies of Na-diffusion that showed
%through pristine and defected MoS$_2$ where it was shown that 
the diffusion barrier for Na from an Mo-substitutional position is 0.48 eV (extracted from plot using \cite{WebPlotDigitizer}) and the presence of a Mo-vacancy defect decreases the diffusion barrier for diffusion through the MoS$_2$ layer (1.470 eV) by almost an order of magnitude compared to pristine MoS$_2$ (14.31 eV)\cite{Yao2019-sd}. 
In our simulations, this process is continuous and leads to the formation of localized regions where continuous flow of dopants through the MoS$_2$ layers (inset to Figure \ref{fig:diffusion bridge}).

The last group of dopants is the non-metals. This group consists of C, Cl, F, N, O, and Si. 
All members of this group had a high formation energy or local structure error.
So the results shown for this group should be considered exploratory and qualitative.
In these simulations, the dopants react with the MoS$_2$ to form chemical compounds. The specific molecular species varies with dopant chemistry. In oxygen-doped systems, we observe oxidation of both Mo and S atoms, to form MoO$_3$ and gaseous SO$_2$ molecules within the simulation box. Carbon dopants form extended chain structures that create interlayer linkages between MoS$_2$ layers along with gaseous CS$_2$. Chlorine and fluorine dopants lead to the formation of molybdenum and sulfur halides, indicating strong halogenation reactions with the host lattice. The simulated behavior of this group of dopants is consistent with previous reports that non-metal dopants are more reactive than the metal dopants and form various chemical compounds with MoS$_2$\cite{Hudec2021-ad,Kotsun2023-yk,Rahman2021-sh, Zhang2017-ek, Farigliano2023-ym}. 
As a representative case from this group, we analyze the N-doped MoS$_2$ simulation, which forms Mo-S-N complexes.

\begin{figure}[ht]
    \centering
    \includegraphics[width=\textwidth]{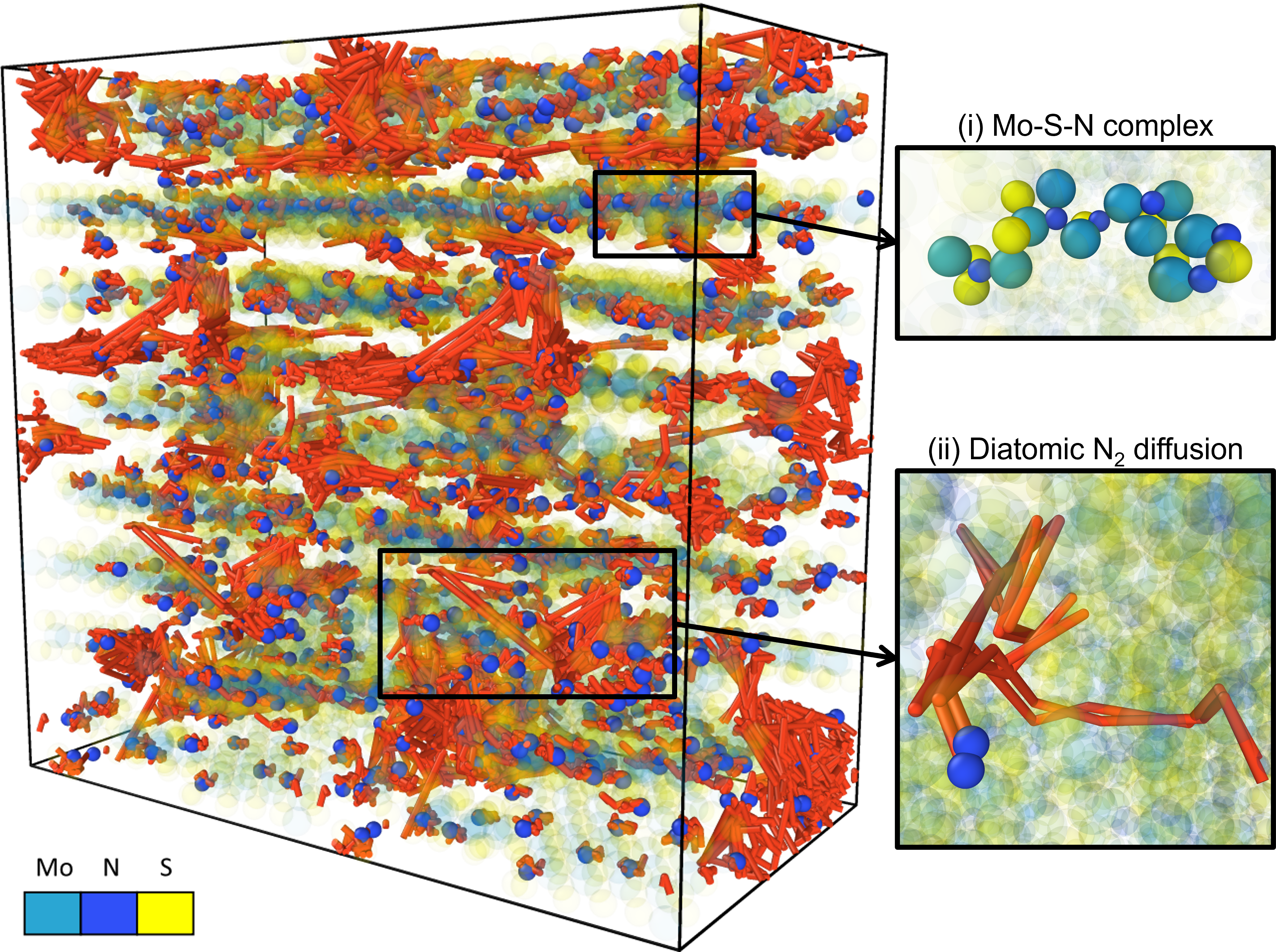}
    \caption{Snapshots from the N-doped MoS$_2$ MLIP simulation with Mo and S made transparent. Most N atoms formed Mo-S-N complexes represented in insert (i) (where Mo and S are made opaque for visualizing the complex) and then remained within the MoS$_2$ lattice. However, some of the N atoms formed gaseous N$_2$ which continued to diffuse throughout the simulation which is represented in (ii).}
    \label{fig:N diffusion}
\end{figure}
\FloatBarrier

Analysis of the atom trajectories from the N-doped MoS$_2$ simulation shows two behaviors. First, many of N dopant atoms chemically react with the MoS$_2$ to form Mo-S-N complexes, as shown in Figure \ref{fig:N diffusion}(i).
Both substitutional and intercalated dopants exhibit this behavior and, once the complexes form, they are very stable such that the dopant atoms have very low diffusivity. 
However, some N atoms exhibits long diffusion paths, as shown in Figure \ref{fig:N diffusion}(ii). 
All instances of this long-range motion occur in pairs where two N atoms move together as a unit. This coordinated movement indicates the formation of N$_2$ molecules within the MoS$_2$ matrix, demonstrating the tendency of nitrogen to maintain its diatomic molecular character even when incorporated as a dopant. 
This shows that the MLIP simulations are effectively able to capture the chemical bond formation and compound synthesis characteristic of MoS$_2$ with non-metal dopants.

\section{\label{sec:conclusions}Conclusions}

This study demonstrates the first application of universal machine learning interatomic potentials for doped MoS$_2$ systems spanning the periodic table. 
By comparing the MLIP-predicted formation energy and local structure with calculations from DFT, we evaluated the accuracy and reliability of the UMA potentials for doped MoS$_2$ systems. Our results show that UMA-small has a mean absolute errors of 0.374 eV and UMA-medium 0.404 eV in formation energy predictions compared to DFT reference calculations. Additionally, through local structure analysis, we showed that the UMA small MLIP is able to capture the lattice distortions caused by the dopants for most cases. This confirmed that the models accurately capture the structural effects of doping across many chemical elements. Our tests also highlighted the limitations of the UMA dataset and identified cases where the MLIP could be finetuned for improved accuracy.

To demonstrate the MLIP, we ran simulations of doped MoS$_2$ systems with $\sim$3,100-atom using UMA small. The heating-cooling molecular dynamics simulations revealed four distinct dopant behaviors in MoS$_2$: clustering metals that aggregated during thermal treatment and could induce layer fracturing, non-clustering metals that maintained mobility without aggregation, light diffusive metals that exhibited through-layer diffusion creating continuous dopant flow channels, and chemically reactive non-metals that form stable molecular compounds within the MoS$_2$ matrix. Analysis of representative examples from these groups showed that the simulation could capture complex phenomena including dopant clustering, interlayer diffusion, chemical compound formation, and structural modifications. The findings provide fundamental insight into dopant-host interactions that govern the performance of doped MoS$_2$ in tribological, electronic, and optoelectronic applications.

Future work will extend the computational framework to investigate dopant concentration effects, temperature dependent phase behavior, and the influence of external force on doped MoS$_2$, as well as to include other doped TMDs.

\section*{Acknowledgements}

The authors thank Daniel Miliate for his technical assistance and review. Dr. Fabrice Roncoroni and Dr. David Prendergast from LBNL for technical assistance. Dr. N. Scott Bobbitt, Dr. Michael Chandross and Dr. John F Curry from SNL for providing feedback on the work.

\section*{Supporting information}

Supplementary\_Information.pdf: MSD calculation methods; MAE based on dopant type (metal or non-metal); Formation energy calculated from DFT and MLMD; Individual dopant MSD and undoped system after heating-cooling cycle.

%%%%%%%%%%%%%%%%%%%%%%%%%%%%%%%%%%%%%%%%%%%%%%%%%%%%%%%%%%%%%%%%%%%%%
%% If you are using classical BibTeX rather than biblatex,
%% remove the \printbibliography and uncomment the \bibliograpy one
%%%%%%%%%%%%%%%%%%%%%%%%%%%%%%%%%%%%%%%%%%%%%%%%%%%%%%%%%%%%%%%%%%%%%
\printbibliography
%\bibliography{acs-template.bib}

\end{document}